%% file: whatIsNP.tex
\documentclass{jicspack}
\input{config}

\usepackage{enumerate}
\usepackage{graphicx}
\usepackage{amssymb}

\begin{document}

\begin{premaker}

% The premaker environment contains Title, authors and addresses;
% \corauth[cor1]{Corresponding author}

\title{What is $NP$? \\
- Interpretation of  a Chinese  paradox  \textit{white horse is not horse}}
\author{Yu LI}
\ead{yu.li@u-picardie.fr}
\address{(1) MIS, Universit\'{e} de Picardie Jules Verne, 33 rue Saint-Leu, 80090 Amiens, France \\
(2) Institut of computational theory and application, Huazhong University of Science and Technology, Wuhan, China}

%\address[author2]{Address for 2nd author}
%\thanks[label3]{Footnotes for address}
% you can use optional labels to link authors explicitly to addresses:
% \author[label1]{},
% \author[label2]{}

\begin{abstract}
% Text of abstract

The notion of \textit{nondeterminism} has disappeared from the current definition of $NP$, which has led to ambiguities in understanding  $NP$, and caused  fundamental difficulties in studying the relation    \textit{P versus NP}. In this paper, we  question the equivalence of  the two definitions of $NP$,  the one  defining $NP$ as the class of problems solvable by a nondeterministic Turing machine in polynomial time, and the other  defining $NP$ as the class of problems verifiable by a deterministic Turing machine in polynomial time, and  reveal  cognitive biases in this equivalence.  Inspired from a famous  Chinese  paradox  \textit{white horse is not horse}, we  further analyze these cognitive biases. The work shows that these cognitive biases arise from   the   confusion between different  levels of  \textit{nondeterminism} and  \textit{determinism}, due to  the lack of understanding about the essence of  \textit{nondeterminism}.  Therefore, we argue that   fundamental difficulties in understanding \textit{P versus NP}  lie firstly  at cognition level, then logic level.

\end{abstract}
\begin{keyword}
    recognition of problem  \sep   $P$ versus $NP$     \sep nondeterminism   \sep determinism   \sep polynomial time verifiability   \sep nondeterministic Turing machine \sep deterministic Turing machine \sep \textit{white horse is not horse}  
\end{keyword}

\end{premaker}

% main text
\section{Introduction}

One of the main issues in the theory of $NP$-completeness is to study whether a $NP$ problem can be efficiently solved by an algorithm, i.e., the existence of a polynomial time algorithm to solve a $NP$ problem. 

However, not only does the general public find it difficult to understand the  theory of  $NP$-completeness, but also a large number of computer scientists and students feel it  half-comprehended. In our opinion, a major cause for this phenomenon is that there exist ambiguities in understanding  $NP$. Initially,  $NP$ refers to  \textit{Nondeterministic Polynomial time}, but the notion of \textit{nondeterminism} is lost in the current definition of $NP$. Consequently, although one can give out the definition of $NP$  as in the literature, but one cannot give a satisfactory explanation about what $NP$ is.  Furthermore,    \textit{P versus NP}   has always remained one of the most perplexing problems in the theory of computational complexity since it was raised in 1971 \cite{cook1}.

In this paper,  we investigate the recognition of $NP$ by questioning the  equivalence of the two definitions of $NP$. 

The paper is organized as follows. In Section 2, we carry out a brief survey of  the definition of $NP$.  In Section 3, we make a preliminary discussion about the recognition of $problem$ with the help of  the etymology in English and in Chinese. In Section 4, we interpret two well-known arguments for supporting  the equivalence of the two definitions of $NP$. In Section 5, we  make use of a famous Chinese paradox \textit{white horse is not horse} to further analyze the cognitive biases in this equivalence. In Section 6, we conclude the paper.

\section{Overview of the  definition of   $NP$}

Concerning   the definition of $NP$,  it can be traced to  the 60's  \cite{lance}\cite{garey}, where  a great number of  applicable and significant problems  were discovered for which no polynomial algorithms could be found to solve them. As time passed and much effort was expended on attempts at efficiently solving these problems, it began to be suspected that there was no such solution.  

Then, the theory of computational complexity  was born  in order to study these problems. The class of  these problems  was defined as problems solvable by nondeterministic Turing machine in polynomial time, denoted as   $NP$ (\textit{Nondeterministic Polynomial time}), to distinguish from those known problems solvable by deterministic Turing machine  in polynomial time, denoted as $P$ (\textit{Polynomial time}). 

Cook's theorem,   presented  in a paper entitled \textit{The complexity of theorem proving procedures} \cite{cook1} in 1971, laid  the foundation for the theory of NP-completeness that becomes the core of the current theory of computational complexity. Since then,  there exist two definitions for $NP$:

\begin{itemize}
\item  \textbf{Solvability-based definition:}
$NP$ is the class of problems solvable  by a nondeterministic Turing machine in polynomial time. 

\item  \textbf{Verifiability-based definition:}
$NP$ is the class of problems verifiable by a deterministic Turing machine in polynomial time. 

\end{itemize}

In order to facilitate the following discussion, we denote the class defined by the verifiability-based definition as $P_{verifiable}$, and keep $NP$ to denote  the class  defined by the solvability-based definition.

The current academic community generally considers the two above  definitions  to be equivalent  \cite{np}, that is, $NP = P_{verifiable}$:

 \textit{The two definitions of NP as the class of problems solvable by a nondeterministic Turing machine in polynomial time and the class of problems verifiable by a deterministic Turing machine in polynomial time are equivalent. The proof is described by many textbooks, for example Sipser's Introduction to the Theory of Computation, section 7.3 \cite{sipser}}.\\

Therefore,  the verifiability-based definition is used  to replace the solvability-based one, and  has become the standard definition of $NP$: 

$P$ is the class of problems  solvable by a deterministic Turing machine in polynomial time, while $NP$ is the class of problems verifiable by a deterministic Turing machine in polynomial time.

Clearly, $P \subseteq NP$, since for each problem in $P$ its solutions can be verified in polynomial time by a deterministic Turing machine. 

Furthermore,   the famous  problem of  \textit{P versus NP}  consists in asking whether a problem verifiable by a polynomial algorithm is   solvable by a polynomial algorithm, denoted as $NP = P$?\\

Such a  problem has become a major unsolved problem in the theory of computational complexity  \cite{garey}\cite{martin}\cite{huang}\cite{scott}, and  was selected as one of the seven Millennium Prize Problems by the Clay Mathematics Institute \cite{cook2}.  Although  a lot of effort has been done (see a comprehensive list maintained by Woeginger (2010) \cite{pnp}), no substantial progress has been achieved until now. 

William Gasarch conducted two polls   about the future of  \textit{P  versus NP}   \cite{william},  with  100 participants in 2002 and 152 participants in  2012. The result (see Table 1) shows that:

\begin{table}
\begin{center}
\begin{tabular}{|c|c|c|c|c|c|c|c|c|c|}
\hline
 & {\it $P  \neq NP$} & {\it $P=NP$} &  {\it Ind}  & {\it DC} & {\it DK} &  {\it DK and DC} &  {\it Other}  \\
\hline
 2002         & $61 (61 \%)$  &  $9 (9 \%)$    & $4  (4\%)$  & $1 (1\%)$  & $ 22 (22\%)$    &  $ 0 (0 \%)$  & $ 3 (3\%)$     \\

2012        & $126 (83\%)$  &  $12 (9\%)$    & $5 (3\%)$  & $5 (3\%)$  & $1 (0.6\%)$    &  $ 1 (0.6\%)$  & $ 1 (0.6\%)$     \\
\hline
\end{tabular}
\end{center}

\caption{Result of two polls about the opinions about \textit{P  versus NP}, where \textit{Ind} stands for Independent of ZFC. \textit{DC} stands for Do not Care. \textit{DK} stands for Do not Know. }
\end{table}
%\vspace{-0.8cm}

\textit{There is a definite trend - more people think $P \neq NP$ now than they did in 2002. How strongly
held are these opinions? Although I did not ask people what the strength of their opinion was in either poll (1) in 2002, 7 out of the 61 $P \neq NP$ votes (11\%) said they had some doubts that $P  \neq NP$, (2) in 2011, 16 out of the 125 (again 11\%) said they had some doubts that $P \neq NP$. Hence, of the people that think $P  \neq NP$, the level of confidence is about the same.} \\

In our opinion, the difficulty  in understanding   \textit{P  versus NP} lies   firstly at cognition level, then  logic level. In fact, due to the fact that  the verifiability-based definition has been accepted as the standard definition of  $NP$, the notion of \textit{nondeterminism} has disappeared from $NP$, which has caused ambiguities in understanding $NP$. So  if we hope to get an insight into    \textit{P versus NP}, we need at first to question the equivalence of the  two definitions of $NP$.  

In the following, we  start from a preliminary discussion about the recognition of problem,  then we   interpret two well-known arguments for  supporting  $NP = P_{verifiable}$. 

 \section{Preliminary discussion about the recognition of problem}

\textit{Problem} is the center of thought. Around a problem, all ideas flourish. So without a problem, there would be no thought.  However,  if one wants to ask some general questions about \textit{problem} such as,    \textit{where does  a problem come from? what is the essence of a problem? how to define a problem?} one would find it difficult to answer these questions, because such an investigation is absent from the traditional philosophy.  \\

Starting from the etymology of \textit{problem} in English and in Chinese, we make a preliminary discussion about the recognition of problem.

\subsection{Etymology of \textit{problem} }

\subsubsection{In English}

In Greek, proballo = pro (in face of) + ballo (to throw). 

\textit{Problem}  means an obstacle to prevent from reaching a goal.  

\subsubsection{In Chinese}

In Chinese,  \textit{problem} consists of two characters, and each character represents an image that is composed by several basic characters  \cite{xushen}\cite{zhu}:

\includegraphics[scale=0.8]{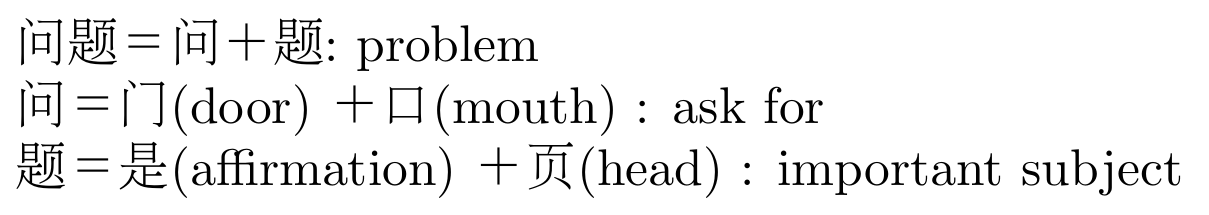}

Then, the Chinese characters of \textit{problem}  describe a scene where  a person outside a door knocks and  asks for a question, and a person inside the door opens and   answers the question.  

Therefore, \textit{problem}  refers to an interface between the questioner and the responder, which means questions about the essence of things.

\subsubsection{Interpretation of \textit{problem}  in the context  of \textit{problem-solving}}

In the context of  \textit{problem-solving}, \textit{problem} can be  interpreted  in two ways inspired by the  above etymology.

When a problem has been recognized out,  it  can be clearly defined by specifying the input  data and  the output result, and solved by designing an algorithm. In this case, such a problem is not really problematique.

When a problem has not been recognized out,  it   is obscure, so it needs  to question  about the essence of things in order to recognize it. In this case, such a problem becomes problematique. 

The two interpretations are complementary and  both contribute to  \textit{problem-solving}.

\subsection{Recognition of   $NP$ }

From the view of  cognition, \textit{recognition} is \textit{cognition} again, that is,  \textit{recognition}   means to distingush something new from  some previously known things, rather than identify it as  those known things.

Such a principe of recognition can be  stated as  \textit{Yin-Yang principe} in the traditional Chinese thought. We refer to a text from \textit{Dao De Jing}, a Chinese classic, to clarify this idea \cite{laozi}: 

\textit{When the people of the world all know beauty as beauty, there arises the recognition of ugliness.  }

\textit{When they all know the good as good, there arises the recognition of bad. }

\textit{Therefore being and non-being produce each other; difficult and easy complete each other; long and short contrast each other; }

\textit{High and low distinguish each other; sound and voice harmonize with each other; beginning and end follow each other.
}  \newline

Concerning   the recognition of $NP$, the key  is to determine the   property that can  distinguish $NP$  from  $P$. 
 
The concept of  $NP$  (\textit{Nondeterministic Polynomial time}) is defined in terms of  nondeterministic Turing machine, that is, $NP$ is  defined at  \textit{algorithm} level rather than  \textit{problem} level. $NP$  implies the  two properties,  the  \textit{nondeterminism}  referring to the multiple choices of  transition, and the   \textit{polynomial time verifiability} referring to  the verifiability of  solution   in polynomial time. 

About the concept of \textit{nondeterminism}, it is under  investigation, so the definition at algorithm level is different from that  at  problem level in the view of cognition, as  discussed in the  above two interpretations of  \textit{problem}. Therefore,   the \textit{nondeterminism} in terms of the multiple choices of  transition  just expresses the formal sense of nondeterminism at algorithm level, but not the essential sense of  nondeterminism at  problem level. In order to continue our discussion, we temporarily use such  nondeterminism as denotation of concept  to recognize $NP$ from $P$. For the concept of \textit{polynomial time verifiability}, it concerns the \textit{determinism} that is already clear in the view of cognition.

The equivalence of the two definitions of $NP$  suggests that    the property of \textit{Polynomial time verifiability}  is used to recognize $NP$.  Let us interpret two well-known arguments for supporting this equivalence. 

\section{Interpretation of two  well-known arguments for  $NP = P_{verifiable}$}

\subsection{A popular argument for  $NP = P_{verifiable}$}

A popular  argument   accepted by the general public can be stated as: 

\textit{Because  the solution of a $NP$ problem  is verifiable in polynomial time,   so $NP$ is the class of problems verifiable in polynomial time ($NP = P_{verifiable}$).}  \\

The argument consists of two basic propositions. The first, \textit{the solution of a $NP$ problem  is verifiable in polynomial time},    means that a   $NP$ problem  is a member of the set $P_{verifiable}$ and possesses the property of polynomial time verifiability.  The second, \textit{$NP$ is the class of problems verifiable in polynomial time},   means   that   the set $NP$ is  identified as the set $P_{verifiable}$, i.e.,  $NP = P_{verifiable}$. The argument uses the  polynomial time verifiability that defines $P_{verifiable}$  to recognize  $NP$ as $P_{verifiable}$. \\

 Although a $NP$ problem  is   verifiable in polynomial time at logic level, but  this property  is not able to recognize $NP$ from $P$ at cognition level, because the essence of this property is the  \textit{determinism} and  it is  shared by  $P$, since a  $P$  problem is  verifiable in polynomial time. Therefore,  this argument is not valid.  In other words,  if  the  polynomial time verifiability is used to define $NP$,  the   \textit{nondeterminism} would disappear at cognition level,  as the \textit{determinism} and the \textit{nondeterminism}  are relative at the same level of concept.

\subsection{A professional argument for  $NP = P_{verifiable}$ }

A professional argument  accepted  in the theory of $NP$-completeness can be stated as:  

\textit{Because a deterministic Turing machine can be considered as a special case of a nondeterministic Turing machine, so  $P \subseteq NP$ thus $NP = P_{verifiable}$}.  \\

In fact, the first proposition, \textit{a deterministic Turing machine can be considered as a special case of a nondeterministic Turing machine},  is  a  pure mathematical deduction, that is, \textit{if  ($n=1$)  and  ($n > 1$) then  ($n \geq 1$)}, where  ($n=1$)  refers to  only one transition,    ($n > 1$)     several possible transitions, and ($n\geq 1$)  one or several possible transitions.  In other words, this proposition is just  valid at mathematical logic  level, because the sense of   the transition number $n$   is not taken into account. For the second proposition, it concerns the recognition of $NP$ and  is situated at cognition level.  \\

Therefore, although  ($n=1$)  and ($n > 1$)  can be mixed to be  ($n\geq 1$) at mathematical  logic level, but  $P$ and $NP$ cannot be confused  in terms of $P \subseteq NP$ at cognition level, because the relation  $P \subseteq NP$  itself needs  to be proved. This \textit{question begging} leads to $NP = P_{verifiable}$, then the nondeterminism disappears from $NP$.

\section{Interpretation of a  Chinese paradox \textit{white horse is not horse}}

The cognitive biases in the above  recognition  of $NP$ are quite  typical and widespread. For example, a famous Chinese paradox  called \textit{white horse is not horse} \cite{horse}, illustrates  such  cognitive biases. 

The  paradox of \textit{white horse is not horse} was  proposed by a great Chinese logician \textit{Gongsun Long} (325 - 250 BC), and it  tells  a history that, one day, \textit{Gongsun Long}  went to a city on riding a white horse. At the gate of the city, the guard said to him that, \textit{your white horse is horse}, in according to regulations,  horses were not allowed to enter the city. Thus, \textit{Gongsun Long}  began his argument - \textit{white horse is not horse}, finally he persuaded the guard, and entered the city riding on his white horse. \\

The argument of \textit{Gongsun Long} was presented in his writing known as  \textit{White Horse Dialogue} between two speakers. Here is a partial summary:

\textit{The issue : Can it be that white horse is not horse?}

\textit{Advocate: It can.}

 \textit{Objector: How so?}
 
\textit{Advocate: HorseÓ is named by means of  the shape. ÒWhiteÓ is named by means of the color. What names the color is not what names the shape. Hence, I say that ÒWhite horse is not HorseÓ.}

\textit{Objector: Having a white horse cannot be said as having no horses. As it cannot be said as having no horses,  it means having a horse. Having a white horse is having a horse, how can a white one not be horse?}

 \textit{Advocate: If one wants a horse, that extends to yellow or black horses. But if one wants a white horse, that does not extend to yellow or black horses. Suppose that a white horse were  horse, then what one wants would be horse. If what one wants were horse, then a white horse would not differ from a horse. If what one wants does not differ, then how is it that a yellow or black horse is sometimes acceptable and sometimes unacceptable? It is clear that acceptable and unacceptable are mutually contrary. Hence, given a yellow or black horse, one can respond that there are horses, but one cannot respond that there are white horses. Thus, it is evident that white horse is not horse.
 }  \newline

We analyze this history.  For the guard, his responsibility was to recognize horses and  to prevent any horse from entering the city, whether it was a white, yellow or black horse.  So when he said  that \textit{your white horse is  horse},  in fact he meant that   \textit{the white horse of Gongsun Long is a member of the set of horses}. However, for Gongsun Long, his objective was to recognize the set of white horses, that is, he wanted to distinguish the set of white horses with the set of yellow or black horses, so  the term of \textit{white horse} in his argument  \textit{white horse is not horse} refers to  the set of white horses instead of his white horse.

The proposition  of the guard and that of Gongsun Long are right respectively. But if they are confused, there would arise cognitive biases,  such as saying \textit{because a white horse is  horse,  so white horse is horse}, because the two propositions situate at two different levels.  In the writing of  \textit{White Horse Dialogue}, Gongsun Long described such cognitive bias  as  a bird flying in  water.

Gongsun Long just took advantage of such  cognitive bias that was not perceived by  the guard, and  confuse the guard, so that the guard forgot his position and slided to the position of  Gongsun Long, finally  let the white horse of Gongsun Long  enter the city. \\

If we make an analogy between the paradox  \textit{white horse is not horse} and the issue of the recognition of $NP$,   we can clearly see  their similarity.

\section{Conclusion}

We question  the equivalence of the two definitions of $NP$ and argue that  there exist cognitive biases in this equivalence that arise from the confusion of  levels of  \textit{nondeterminism} and  \textit{determinism}, due to the lack of understanding about the  essence of \textit{nondeterminism}.  Rightly in this sense, we argue that the difficulty in understanding   \textit{P versus NP} lies   firstly  at cognition level, then  logic level. 

This work is the first step of our work  of the interpretion of the theory of $NP$-completeness \cite{li}, and undertaken by another work of the interpretation of Cook's theorem \cite{li2}. Moreover, it  provides an interpretation of a deeper work discussed in \cite{zhou1}\cite{zhou2}. 

We hope that this work can evoke reflections from different angles about some fundamental problems in cognitive science, and contribute to understand  \textit{P  versus NP}. Furthermore, we hope that this work can help to understand the complementarity of Chinese thought and Western philosophy.

\section*{Acknowledgement}

This work has been guided and discussed with  my masters Mr JianMing ZHOU, Mr  BangFu ZHU, Mr  WenQi HUANG and  Mr Jacques CARLIER. It is also  the fruit of the exchange with  my colleagues.

%\section*{Appendix} Appendix here.

\end{document}

%% file: config.tex
\setvolume{8}                             %for printing use only
\setyear{2012}                             %for printing use only
\setpagerange{1}{8}                    %for printing use only
\setheadauthor{X. Liu et al.}          %for printing use only
\setissn{1553--9105} \setpubdate{January 2012} \setno{1}

\afterpage{\beginheader}                   %for printing use only